\documentclass[aps,prl,prev,twocolumn,superscriptaddress,floatfix,nofootinbib]{revtex4-1}
\pdfoutput=1

\usepackage{graphicx}
\usepackage{bm}
\usepackage{times}
\usepackage{slashed}
\usepackage{color}
\usepackage{aas_macros}
\usepackage{slashed}
\usepackage{lipsum}
\usepackage{subfigure}
\usepackage{multirow}
\usepackage{amsmath}
\usepackage{hyperref} 
\usepackage{todonotes}
\usepackage[normalem]{ulem}
\hypersetup{colorlinks = true, 
citecolor=blue,
linkcolor=blue,
filecolor=blue,      
urlcolor=blue
}

\begin{document}

\title{Second order Fermi reacceleration mechanisms and large scale synchrotron radio emission in intra-cluster bridges}
\author{Gianfranco Brunetti}
\email{brunetti@ira.inaf.it}
\affiliation{INAF- Istituto di Radioastronomia, via P. Gobetti 101, Bologna, Italy}

\author{Franco Vazza}
\email{franco.vazza2@unibo.it}

\affiliation{Dipartimento di Fisica e Astronomia, Universit\'{a} di Bologna, Via Gobetti 93/2, 40121, Bologna, Italy}
\affiliation{INAF- Istituto di Radioastronomia, via P. Gobetti 101, Bologna, Italy}
\affiliation{University of Hamburg, Hamburger Sternwarte, Gojenbergsweg 112, 21029 Hamburg, Germany}

\begin{abstract}
Radio observations at low frequencies with the Low Frequency Array (LOFAR) start discovering gigantic radio bridges connecting pairs of massive galaxy clusters. These observations probe unexplored mechanisms of {\it in situ} particle acceleration that operate on volumes of several Mpc$^3$. Numerical simulations suggest that such bridges are dynamically complex and that weak shocks and super-Alfv\'{e}nic turbulence can be driven across the entire volume of these regions. In this Letter we explore, for the first time, the role of second order Fermi mechanisms for the reacceleration of relativistic electrons interacting with turbulence in these peculiar regions.
We assume the turbulent energy flux measured in simulations and adopt a scenario in which relativistic particles scatter with magnetic field lines diffusing in super-Alfv\'enic turbulence and magnetic fields are amplified by the same turbulence. We show that steep spectrum and volume filling synchrotron emission can be generated in the entire intra-cluster bridge region thus providing a natural explanation for radio bridges. Consequently, radio observations have the potential to probe the dissipation of energy on scales larger than galaxy clusters and second order Fermi mechanisms operating in physical regimes that are still poorly explored. This has a potential impact on several branches of astrophysics and cosmology.
\end{abstract}

\maketitle

{\bf Introduction -- } Mpc-scale, steep-spectrum, diffuse radio emission from the intra-cluster-medium (ICM) is observed in the form of giant radio halos and relics in dynamically active and massive galaxy clusters \citep[e.g.][for reviews]{2019SSRv..215...16V, bj14}. 
This suggests that part of the kinetic energy that is dissipated in the ICM during cluster-cluster mergers is channelled into the acceleration of relativistic particles and amplification of magnetic fields through a hierarchy of complex mechanisms that transfer energy from Mpc scales to small scales, and that presumably involve shocks and turbulence operating in a unique plasma regime \citep[e.g.][]{cassano05,brunettilazarian07,miniati15, brunetti16, xuspitkovskycaprioli19, ryu19, kangryu11, kangryu12, ry08, miniatinature15}.\\
More recently LOFAR observations have discovered diffuse radio emission
from regions extending on even larger scales and that connect pairs of massive clusters in a pre-merger phase \citep[][]{2018MNRAS.478..885B,2019Sci...364..981G}. 
These observations prove that these regions, where the gas is likely compressed, are filled by relativistic electrons and magnetic fields that are generated on scales which had never been probed before.
The most spectacular case is the 5 Mpc long radio bridge connecting the two massive clusters A399 and A401 \citep[][]{2019Sci...364..981G}, where the radio emission follows a filament of gas connecting the two clusters that was early discovered with the Planck satellite through the Sunyaev-Zeldovich effect \citep[][]{2013A&A...550A.134P}.\\
What makes their interpretation challenging is that radio bridges appears as truly diffuse radio emissions on gigantic scales, suggesting that relativistic particles are accelerated {\it in situ} by mechanisms that are distributed on very large spatial scales and that are not necessarily powered by the energy dissipated as a consequence of major cluster-cluster mergers.
While recent numerical simulations have suggested that {\it equatorial} shocks can be launched perpendicular to the merger axis even in a pre-merger phase \citep[e.g.][]{2018ApJ...857...26H}, strong shocks are very rare in the hot and compressed gas of intra-cluster bridges \citep[e.g.][]{vazza19}. For this reason the large area filling factor observed in the radio bridge of A399-A401 clearly disfavours shock acceleration as the main source of the observed emission and suggests that pre-existing and volume filling supra-thermal electrons are (re)accelerated to radio-bright energies ($>$ GeV) by other mechanisms.
Numerical simulations show that relatively weak shocks, $\mathcal{M} \leq 2-3$, form in these regions and that up to $\sim 10\%$ of the volume has been crossed by these shocks in the last Gyr \citep[e.g.][]{2019Sci...364..981G} leading to the possibility that
radio bridges may result from re-acceleration of a volume filling population of {\it fossil} relativistic electrons by these weak shocks under favorable projection effects. However, in order to match the constraints on the spectrum of the emission, this scenario requires assumptions on the age and dynamics of the {\it fossil} electrons that are not very plausible (see discussion in \citep[][]{2019Sci...364..981G}).
Virtually all major mergers should undergo a stage in which the remnant of a cosmic filament connecting the two clusters is compressed and pre-processed by gas dynamics, before the two clusters collide. 
Therefore, recent detections \citep[i.e.][]{2018MNRAS.478..885B, 2019Sci...364..981G} may have unveiled the tip of the iceberg of a common (albeit short-lived, i.e. $\sim \rm ~Gyr$) phenomenology.
Understanding the mechanisms of acceleration of radio emitting particles in such pillars of the cosmic web is therefore also key to prepare to what the future generation of radio surveys will deliver \citep[e.g.][]{vazza19}.
In this Letter, we propose that fossil electrons, released in the ICM in the past by the activity of AGN and star-forming galaxies, are re-accelerated by the turbulence in the regions bridging massive pre-merging systems.
\begin{figure}
\centering
\includegraphics[width=0.55\textwidth]{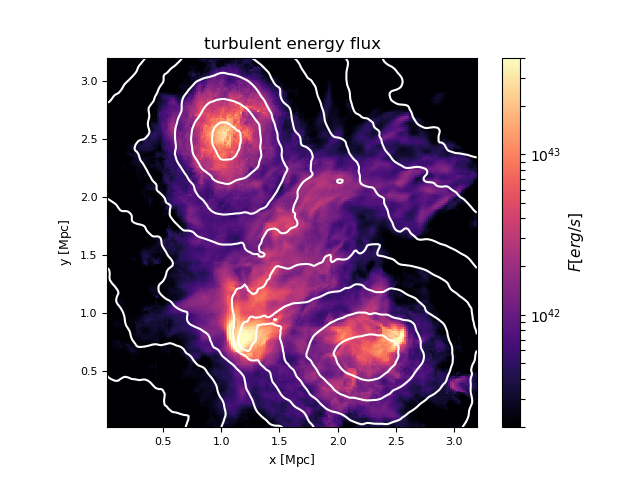}
\caption{Map of kinetic energy flux integrated along the line of sight ($5.1 \rm ~Mpc$) for the simulated system at $z=0.1$, with logarithmically spaced gas projected density contours ($\Delta \rm log_{\rm 10}n=0.25$).}
\label{Fig:maps}
\end{figure}
\\
{\bf Dynamics and turbulence in bridges connecting clusters --} 
Massive binary mergers are rare and powerful events occurring in high over-density regions \citep[e.g.][]{1997ApJS..109..307R}. 
Even during its early stage, the dynamics of the collapse and the accretion of smaller sub-clusters drive weak shocks \citep[e.g.][]{2018ApJ...857...26H,vazza19} and transonic turbulence \citep[e.g.][]{iapichino11}.
We used cosmological MHD numerical simulations obtained with the ENZO code \citep[][]{enzo14} to examine 
the properties of turbulence and magnetic fields in 
a binary cluster collision during its pre-merger phase.
Specifically, we used the same pair of simulated clusters presented in \citep[][]{2019MNRAS.486..623D, 2019Sci...364..981G}, that closely resemble the A399-A401 pair which is the reference of our work.
Simulations have 8 levels of Adaptive Mesh Refinement (AMR) to selectively increase the spatial and force resolution down to $\Delta x=3.95 \rm ~ kpc/cell$ in most of the virial volumes of clusters, and assume a uniform primordial magnetic field $B_0=10^{-10} \rm ~G$ (comoving) at the start of the simulation.
\\
We attempt to separate turbulent fluctuations, $\delta V$, from bulk motions on larger scales using the filtering technique in \citep[][]{va17turb} and extract solenoidal, $\nabla \cdot \vec{v}=0$, and compressive, $\nabla \times \vec{v}=0$, turbulent components using the Hodge-Helmholtz projection in Fourier space \citep[e.g.,][]{va17turb}.
This allows to estimate the local turbulent energy flux across scales, $F \sim 1/2 \rho_{ICM} \delta V^3/L$, which is scale-independent in Kolmogorv turbulence.
We measure that $\sim 60\%$ of the turbulent energy flux in the bridge is associated to solenoidal motions.
Figure \ref{Fig:maps} shows the projected energy flux of the solenoidal component measured when the two clusters are in a pre-merger stage ($z=0.1$), in a situation (cluster masses and dynamics) similar to A399-A401. 
Turbulence in the bridge has injection scales $L \sim$ 400 kpc - 1 Mpc \citep[see also][]{vazza19} and is powered by the accretion of matter and smaller sub-clusters (visible in Fig.\ref{Fig:maps} and Supp. Material) on the over-dense region containing the two main clusters; the detection of X-ray bremsstrahlung from these sub-clusters and from the bridge itself is however challenging with present X-ray telescopes \citep[see discussion in Sec. 3.3.4 of][]{vazza19}. In particular, solenoidal motions originate from baroclinic instabilities at curved shocks and compressive amplification of accreted vortical motions \citep[e.g.][]{iapichino11,po15,wi17b}, and from the generation of vorticity by shear stresses \citep[e.g.][]{va17turb}.\\
In the region connecting the clusters (a cylinder 1.5 Mpc $\times$ 3 Mpc, $V \sim 5$ Mpc$^3$) we measure a turbulent luminosity $F \cdot V \sim 10^{45}$erg s$^{-1}$ (masking regions around small sub-clusters where $F$ can be biased high by our filtering). This is similar to the luminosity found in simulated clusters during mergers \citep[e.g.][]{2009MNRAS.399..410P, va17turb}. 
Solenoidal turbulence is a key ingredient for magnetic field amplification in the ICM via small scale dynamo \citep[e.g.][]{dolag05, po15, ry08, miniatinature15}, which indeed is a mechanism that is observed in the central regions of clusters in MHD simulations \citep[][]{zuhone11, va18mhd, 2019MNRAS.486..623D,donnertrev18}.
However, in intra-cluster bridges this process is quenched by the limited spatial resolution in our simulations
$\Delta x \gg l_A$; $l_A$ being the MHD scale where the velocity of turbulent eddies equals the Alfv\'en speed and where most of the amplification takes place. We thus estimate the field in post processing (Supp Material).
The plasma in intra-cluster bridges shares conditions similar to the medium in the outskirts of galaxy clusters, being a weakly collisional and unstable high beta plasma
with presumably very high effective Reynolds number \citep[e.g.,][]{sche06, lazarianberesnyak06,instabilities10,brunettilazarian11mfp,santos14}. Under these conditions, after the turbulent cascade reaches dissipation scales, a fixed fraction of the energy flux of MHD turbulence is channelled into magnetic field \citep[e.g.,][]{beresnyak12}. We thus estimate the magnetic field in our simulation as $B^2/8\pi \sim \eta_B F \tau_e \sim {1 \over 2} \eta_B \rho_{ICM} \delta V^2$, where $\tau_e$ is the eddy turnover-time $\tau_e \sim L/\delta V$, and $\eta_B \sim$ few percent.
We obtain a volume-averaged field in the bridge $<B> \sim 0.5-0.6 \mu$G, that is $\sim 3$ times larger than the original field in our simulations. This is $\sim 3-5$ times smaller than the typical field in the internal regions of galaxy clusters \citep[e.g.][]{2019MNRAS.486..623D, 2019SSRv..215...16V} implying values of the beta-plasma, $\beta_{pl}\sim 100-200$, that are slightly larger than those in clusters. This is because bridges are dynamically younger regions, and their life-time 1-1.5 Gyr (collision time of clusters) is comparable to the turbulent eddy-turnover times, $\tau_e\sim 0.4-1$ Gyr, in these regions. 
\\
{\bf Turbulent reacceleration model --} Turbulent acceleration drains a fraction of the turbulent energy flux into particles :
\begin{equation}
  \frac{\rho_{ICM} \delta V^3} {L} \eta_{_{CRe}} \sim \int d^3p E {{\partial f_e}\over{
\partial t}}
\label{turboflux}
\end{equation}
where $\eta_{_{CRe}}$ is the electron acceleration efficiency, the right term describes the energy flux into accelerated electrons, and $f_e$ is the electrons distribution function in the momentum space.
Radio emitting electrons in the ICM lose energy mainly through synchrotron emission and inverse Compton (IC) scattering off the Cosmic Microwave Background (cmb) photons.
The turbulent luminosity measured in the simulated bridge ($\approx 10^{45}$erg s$^{-1}$) should be compared to the total (IC and synchrotron) non-thermal luminosity of the bridge, $L_{IC+S} \sim L_S (1+(B_{cmb}/B)^2)$, where $L_s \sim 10^{40}$erg s$^{-1}$ is the radio luminosity of A401-A399 \citep[][]{2019Sci...364..981G} and $B_{cmb}=3.25 (1+z)^2 \mu$G. If we assume the magnetic field derived in the previous Section, the turbulent luminosity is $\gg 1000$ times the total non-thermal luminosity from the bridge, and thus only a small fraction of the turbulent energy flux is required to maintain the non-thermal emission.
Energetic particles in a turbulent medium can be subject to second order Fermi acceleration.
Several studies considered the Transit Time Damping (TTD) with compressive modes in the ICM
\citep[e.g.][]{brunettilazarian07,miniati15,pinzke17}.  More recently,  \citet{2016MNRAS.458.2584B} proposed a mechanism that operates in large-scale super-Alfv\'enic solenoidal turbulence in the ICM, where particles are reaccelerated stochastically diffusing across regions of magnetic reconnection and dynamo \citep[see][for application to gamma-ray bursts and Pulsar wind nebulae]{xu17, xu19}. On much smaller scales, situations involving first order and second order Fermi-like acceleration are also observed in simulations of reconnection regions \citep[e.g.,][]{kowal12, dahlin14, guo19, comisso19}.
In the case of prevalence of solenoidal component and strongly
super-Alfv\'enic turbulence, $M_A^2=(\delta V /V_A)^2 \sim M^2 \beta_{pl} \gg 1$, as in the simulated bridges where $M_A \sim 8-10$ ($M_A \sim 30$ assuming the original field values from simulations), this acceleration mechanism may become faster than TTD (Supp Material).
Thus following \citep[][]{2016MNRAS.458.2584B} we adopt a diffusion
coefficient in the particle momentum space (assuming a reference value for the
effective mfp of relativistic electrons $\sim 1/2 l_A$, Supp Material):
\begin{figure}
\centering
\includegraphics[width=0.47\textwidth]{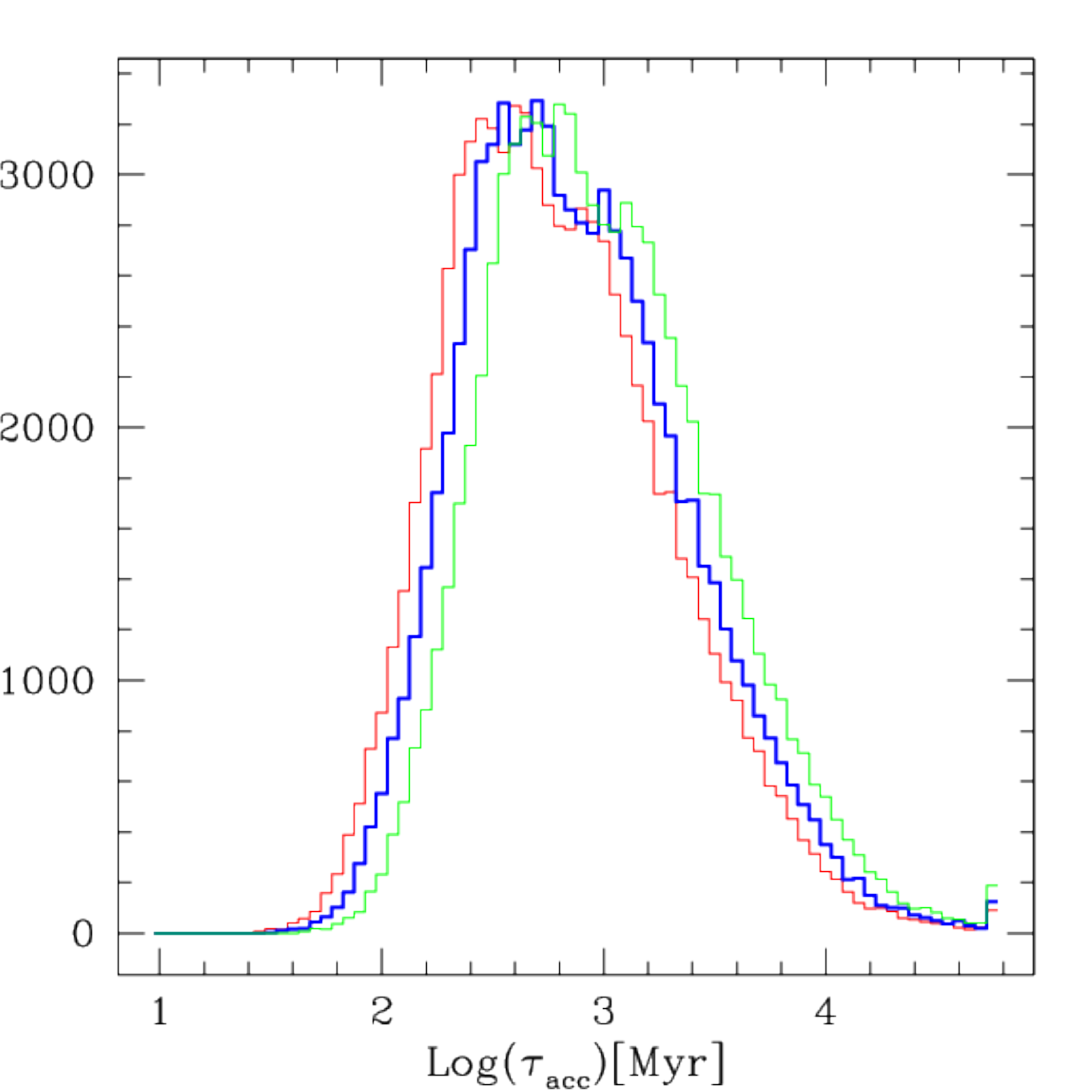}
\caption{Number of cells as a function of the particle acceleration time (81,000 cells sampling the simulated bridge) assuming $\eta_B$=0.02 (red), 0.03 (blue), 0.05 (green).}
\label{Fig:time_acc}
\end{figure}
\begin{figure*}
\centering
\includegraphics[width=0.47\textwidth]{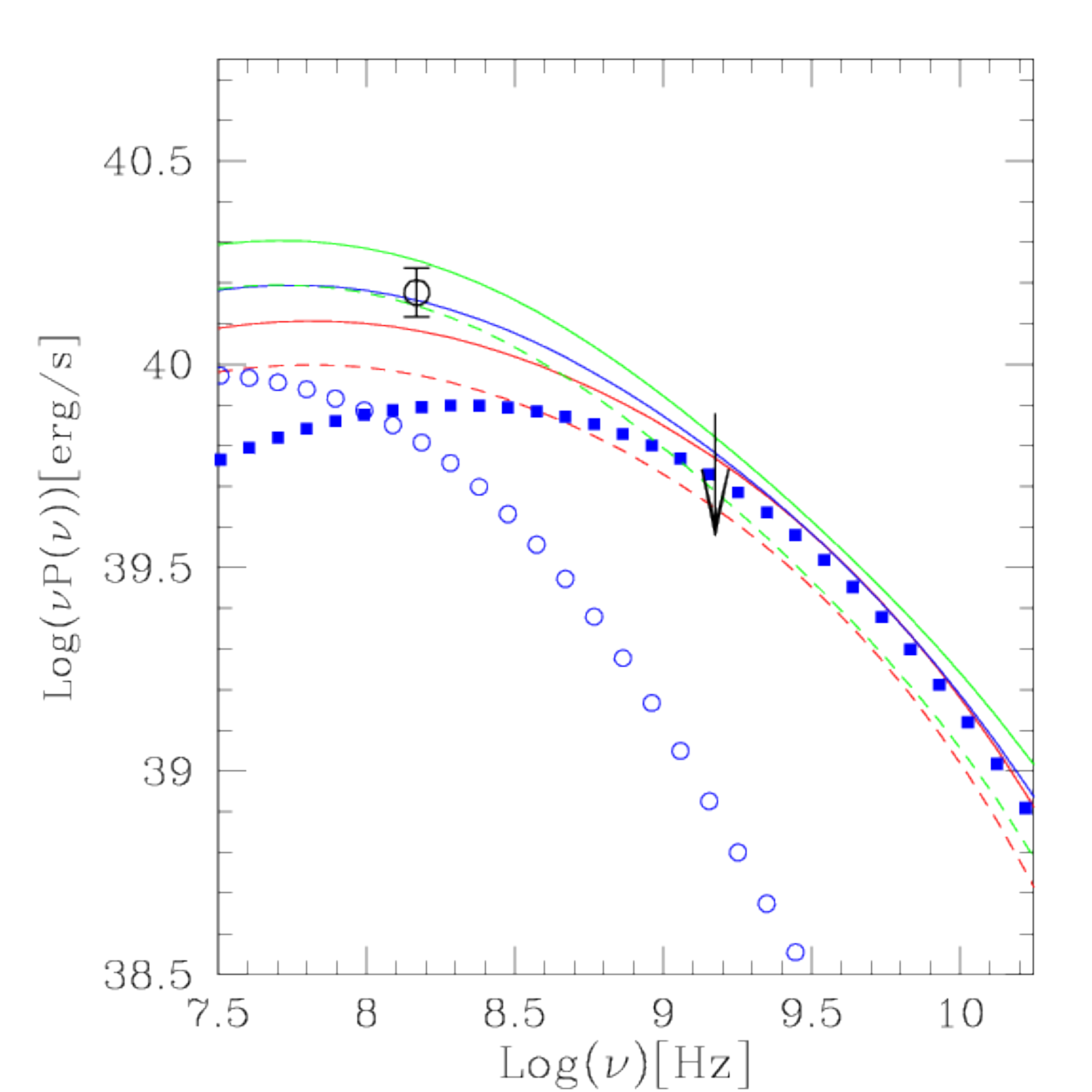}
\includegraphics[width=0.47\textwidth]{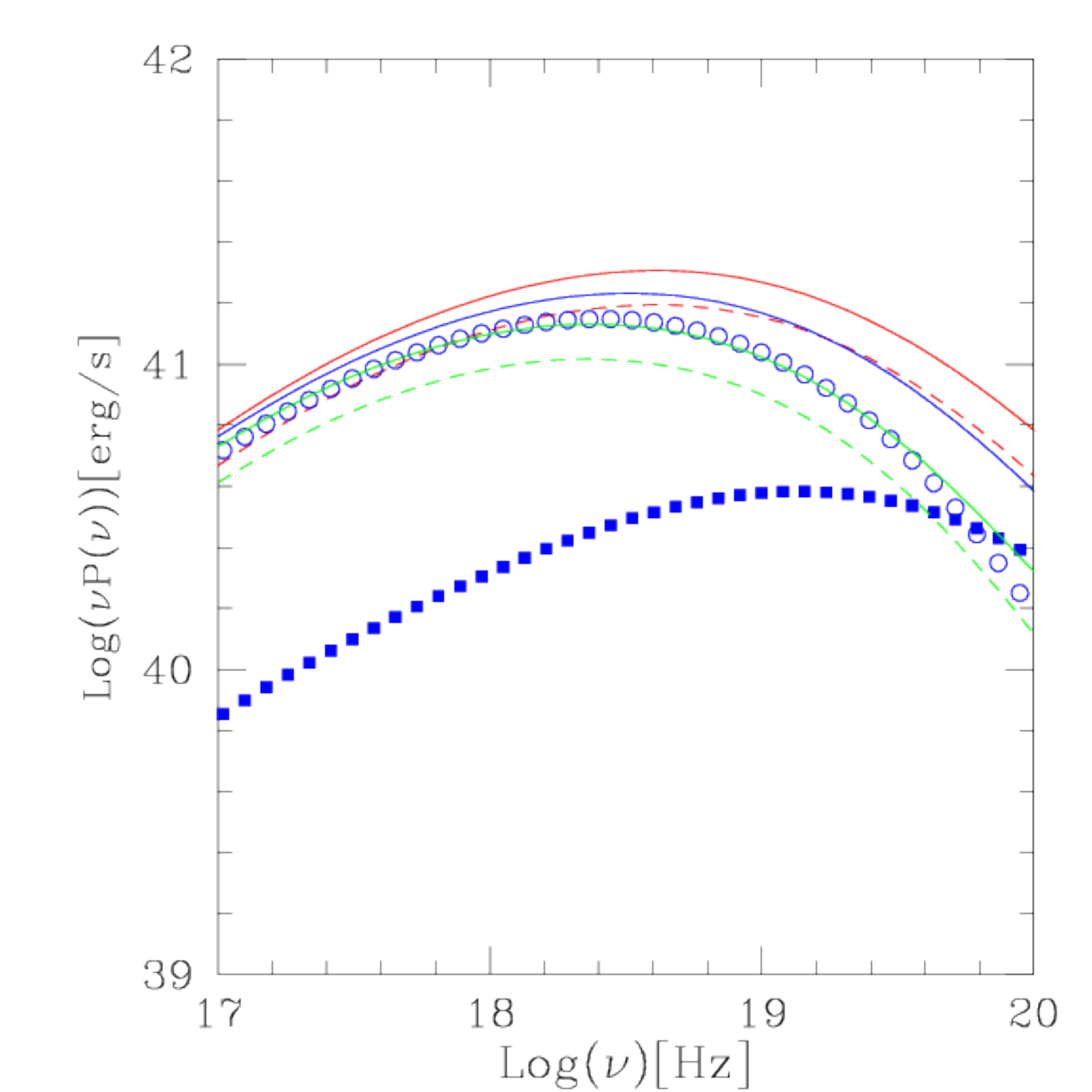}
\caption{Synchrotron (left) and IC (right) spectra obtained for
$\eta_B$=0.02 (red), 0.03 (blue) and 0.05 (green). Dashed lines (for $\eta_B$=0.02 and 0.05) mark models assuming only primary electrons. The contribution from regions with larger (85\% of cells in the volume) and shorter acceleration times (15\% of the volume) are also marked with open and filled points, respectively, considering $\eta_B=$0.03. 
Radio data are taken from \citep[][]{2019Sci...364..981G}.}
\label{Fig:spettri}
\end{figure*}
\begin{equation}
D_{pp} \sim {{48}\over c} {{F}\over{\rho_{ICM} V_A}} p^2
\label{dpppsi}
\end{equation}
By adopting our magnetic field model, the Alfv\'en speed is $V_A \simeq \sqrt{{2 \over {\rho_{ICM}}} {{F L}\over{\delta V}} \eta_B}$ and $D_{pp} \propto p^2 \eta_B^{-1/2}{{\delta V}^2}/L$.
We thus use eq.\ref{dpppsi} and the energy flux of the solenoidal turbulence measured in the simulated bridge to calculate the electrons re-acceleration time $\tau_{acc} = p^2/(4 D_{pp})$. 
We calculate the acceleration time in $81,000$ cells sampling the region connecting the two simulated clusters (Supp Material). The results are shown in Fig.\ref{Fig:time_acc} for three values of $\eta_B$. The acceleration times are very long, as expected for second order Fermi mechanisms, however they are much shorter than the dynamical time-scale of bridges and the turbulent eddy turnover times.
Most important, in $\sim 1/3-1/2$ of the volume the acceleration time is similar to, or smaller than, the cooling time of radio emitting electrons at the redshift of the A399-A401 system, $\tau \sim 220 (B_{\mu G}/0.5)^{1/2} (\nu_{MHz} / 150)^{-1/2}$ Myr (assuming $B_{cmb}^2 \gg B^2$). This allows us to conclude that the mechanism can naturally generate volume filling synchrotron emission from the entire bridge. 
\\
{\bf Spectrum of the emission --} Next we evaluate whether the radio spectrum of the bridge in A399-A401 can be explained by our model.
We calculate the evolution of the electrons distribution function, $N = 4 \pi f p^2$, in the general situation in which relativistic electrons and protons, injected in the volume in the past by galaxies and AGN, coexist. We combine Fokker-Planck equations for primary and secondary electrons:
\begin{eqnarray}
  {{\partial N_e(p,t)}\over{\partial t}} =
  {{\partial}\over{\partial p}}
  \Big(
  N_e(p,t) \Big[
{\cal S}_e(p)
      - {{p}\over {3}} (\nabla \cdot V) \Big] \Big) +
\nonumber\\
{{\partial }\over{\partial p}}
      \Big(
      D_{pp} {{\partial N_e(p,t)}\over{\partial p}} 
     - {2\over p} N_e(p,t) D_{pp}  \Big)
      + Q_e(p,t)
\label{electrons}
\end{eqnarray}
and protons 
\begin{eqnarray}
  {{\partial N_p(p,t)}\over{\partial t}} =
  {{\partial}\over{\partial p}}
  \Big(
  N_p(p,t) \Big[
{\cal S}_p(p)
      - {{p}\over {3}} (\nabla \cdot V) \Big] \Big) +
\nonumber\\
{{\partial }\over{\partial p}}
      \Big(
      D_{pp} {{\partial N_p(p,t)}\over{\partial p}} 
     - {2\over p} N_p(p,t) D_{pp}  \Big)
      - {{N_p(p,t)}\over{\tau_{pp}}}
      \label{protons}
\end{eqnarray}
${\cal S}$ accounts for the energy losses of electrons (Coulomb, ICS and synchrotron) and protons (Coulomb), $D_{pp}$ is given by eq.\ref{dpppsi}, $\tau_{pp} = (n_{ICM} \sigma_{pp} c)^{-1}$ is the timescale of inelastic pp collisions in the ICM, $Q_e$ is the injection spectrum of secondary electrons by pp collisions, and $\nabla \cdot V$ accounts for compression \citep[e.g.,][for details]{brunetti17}.
In principle, the evolution of particles should be computed by following their spatial advection with a Lagrangian tracer approach \citep[e.g.][]{wi17b} and then by integrating in time eqs.\ref{electrons} and \ref{protons} for each tracer \citep[e.g.][]{2014MNRAS.443.3564D}.
However, this approach is numerically challenging and clearly beyond the exploratory goal of the present Letter.
Here we adopt a simple {\it single zone} model, assuming average quantities that are measured in the simulated bridge region at a fixed time $z=0.1$, namely $kT=5$ keV, $n_{ICM}=3\times 10^{-4}$cm$^{-3}$ (both consistent with measurements in A399-A401), and $\nabla \cdot V \sim 0.75\times 10^{-16}$s$^{-1}$.
We then assume different $\tau_{acc}$ spanning the range of values in Fig.\ref{Fig:time_acc} and for each value of $\tau_{acc}$ calculate electrons spectra from eqs.\ref{electrons} and \ref{protons} assuming $B=<B>$ and evolving spectra for one turbulent eddy turnover time $<\tau_e>$. Specifically, for each $\tau_{acc}$, $<B>$ and $<\tau_e>$ are obtained by averaging the values of $B$ and $\tau_e$ in the cells with acceleration time  $=\tau_{acc}$.
Finally, we obtain the emission integrated from the bridge region by combining the emissions generated by each electron spectrum weighted for the probability distribution function of the acceleration times at z=0.1 (from Fig.\ref{Fig:time_acc}). 
The remaining ingredient, is the initial spectrum and number of the {\it seed} electrons and protons to re-accelerate. This is largely unknown in bridges and filaments connecting clusters. However, as in the case of clusters, we expect that {\it seeds} primary electrons injected by the past activity of shocks, AGN and Galactic Winds, can be accumulated in the
entire region of the bridge at energies of $\sim$ 100 MeV where their cooling time is maximised \citep[e.g.,][]{bj14}. 
In Figure \ref{Fig:spettri} we show the synchrotron and IC spectra calculated assuming the volume of the radio bridge in A399-A401 $=$5 Mpc$^3$ \citep[][]{2019Sci...364..981G} for an initial spectrum of primary electrons and protons injected at $z=0.2$ and passively evolved to $z=0.07$; the final results are only little sensitive on the exact shape of the initial spectra as they evolve non-linearly with time due to turbulent acceleration and losses.
The initial energy densities of relativistic protons and primary electrons in Figure \ref{Fig:spettri} (solid lines) are assumed $10^{-2}$ and $3 \cdot 10^{-5}$ of the thermal ICM; these are typical values assumed in radio halo models.
In Figure \ref{Fig:spettri} we also show the case with only primary electrons (dashed lines), i.e. without including protons.
Figure \ref{Fig:spettri} shows that the synchrotron spectrum peaks at few hundred MHz, matching well the LOFAR detection \citep[][]{2019Sci...364..981G}, and extends to higher frequencies, where detections are still missing. The IC spectrum peaks in the hard X-rays with a luminosity $\sim$10-20 times larger than the synchrotron luminosity.
Spectra are sensitive to the turbulent energy flux measured in simulations and scale (linearly) with the amount of seed electrons to reaccelerate, whereas they are not very sensitive to $\eta_B$.
The cut-off synchrotron frequency emitted by the re-accelerated electrons is $\nu_c \propto p_{m}^2 B$, where $p_{m} \sim 4 D_{pp}/{\cal S}$ is the maximum momentum of electrons. In our model (for $B_{cmb}^2 \gg B^2$) this gives:
\begin{equation}
    \nu_c \propto F^2 \rho_{ICM}^{-1} \epsilon_t^{-1} \eta_B^{-{1 \over 2}}
    \label{nu}
\end{equation}
where $\epsilon_t \sim 1/2 \rho_{ICM} \delta V^2$. The cut-off synchrotron frequency depends on the turbulent energy flux and turbulent energy density. Consequently a natural prediction of our model is that the synchrotron emission at lower frequencies should be more volume filling, while at higher frequencies it should be contributed by the most turbulent regions that fill a smaller fraction of the volume. 
This is indeed shown in Figure \ref{Fig:spettri} where we report the synchrotron and IC spectrum from cells with $\tau_{acc} > \tau_*$ (filling 85 \% of the volume) (empty circles) and with $\tau_{acc} < \tau_*$ (15 \% of volume) (filled squares); $\tau_*$ a threshold value. Finally, we notice that weak shocks in the bridge \citep[][]{2019Sci...364..981G} may also compress the population of turbulent re-accelerated electrons and the magnetic fields increasing the radio brightness at their location.
\\
{\bf Conclusions --} 
In this Letter we propose that the radio bridges extending on scales larger than clusters originate from second order Fermi acceleration of electrons interacting with turbulence. We show that the physical conditions and very long dynamical time-scales in bridges connecting clusters allow the effects of these gentle mechanisms to become important.
Turbulence is generated by the complex dynamics of substructures in bridges and thus, according to the proposed scenario, radio observations are also novel probes of the dynamics and dissipation of gravitational energy on very large scales.
More specifically we extract the turbulent properties measured in cosmological simulations mimicking the A399-A401 system and assume second order Fermi mechanism from the interaction of relativistic particles with magnetic field lines diffusing in super-Alfv\'enic turbulence.
We demonstrated that the mechanism allows for the re-acceleration of radio emitting electrons in a large fraction of the volume. This can generate a volume-filling synchrotron emission with luminosities compatible with the observed ones and steep spectra, with $\alpha \sim 1.3-1.5$ between $0.15-1.5$ GHz ($L(\nu) \propto \nu^{-\alpha}$) or steeper. The same turbulence amplifies magnetic fields in the bridge. This results in a field that is stronger than that obtained directly from current cosmological simulations, with a potential impact on studies based on Faraday Rotation and on the propagation of very high energy cosmic rays.
Future observations will test crucial predictions of the model: the filling factor of the radio emission should be larger at the low frequencies observable with LOFAR, making the emission smoother there, while it is predicted to decrease at higher frequencies, where the emission gets dominated by the clumpy contribution from smaller regions with high acceleration rate. 
Finally, our model predicts IC emission in the hard X-rays with a luminosity 10-30 larger than the synchrotron component.

We acknowledge the referees and discussions with R.Cassano, F.Govoni and J.Drake. 
FV acknowledges computing time through the John von Neumann Institute for Computing on the supercomputer JUWELS at J\"{u}lich Supercomputing Centre (projects hhh44 and stressicm), and financial support through the ERC Starting Grant MAGCOW, no.714196.

\bibliography{franco2}

\section{Additional Figures \& projection effects}

For completeness in Figure \ref{Fig:dens} we report the exact position of the 3 sub-boxes (for a total of 81000 cells) used for the analysis of the bridge, superimposed to the gas density distribution.
The two panels in Fig.\ref{Fig:mapXY} show the (solenoidal) turbulent kinetic energy flux projected through the simulation along the two directions perpendicular to the axis used in Fig. 1 (paper): it is clear that the turbulence fills the whole region connecting the two clusters, with local maxima associated with substructures crossing each different line of sight.

\section{MHD scale and magnetic field model}

\begin{figure}
\centering
\includegraphics[width=0.47\textwidth]{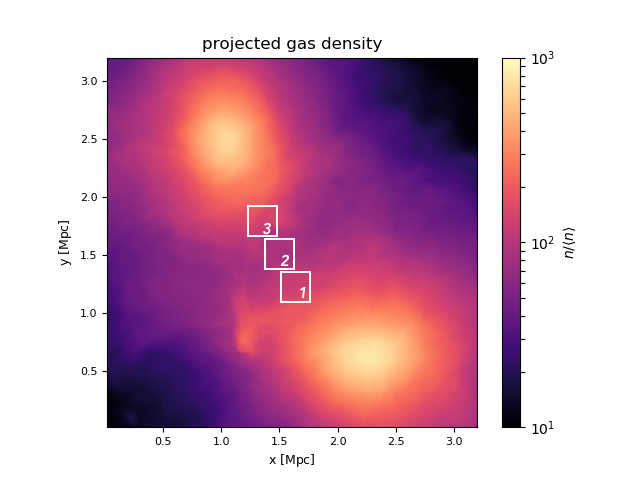}
\caption{Projected gas density for the simulated merger and location of the three $\approx 480^3 \rm ~kpc^3$ sub-volumes used for our analysis.}
\label{Fig:dens}
\end{figure}

\begin{figure}
\centering
\includegraphics[width=0.47\textwidth]{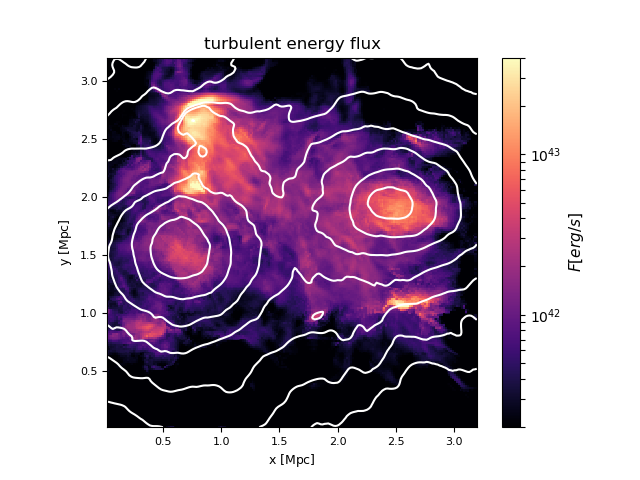}
\includegraphics[width=0.47\textwidth]{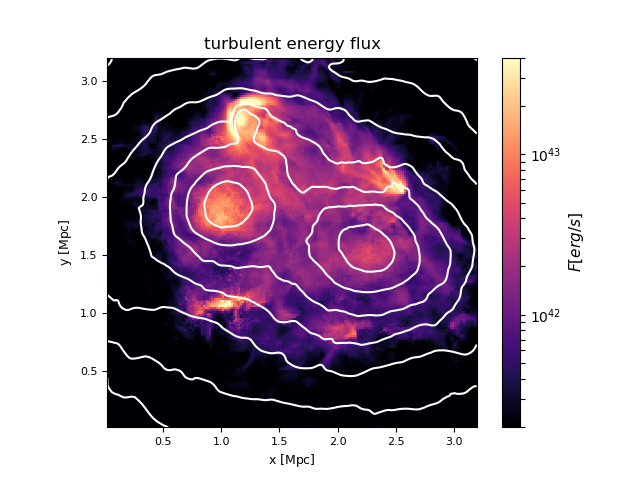}
\caption{Map of integrated 
kinetic energy flux along the line of sight ($5.1 \rm ~Mpc$ for our simulated cluster collision at $z=0.1$) as in Fig.1 (paper), but for two other perpendicular lines of sight.}
\label{Fig:mapXY}
\end{figure}

\begin{figure}
\centering
\includegraphics[width=0.47\textwidth]{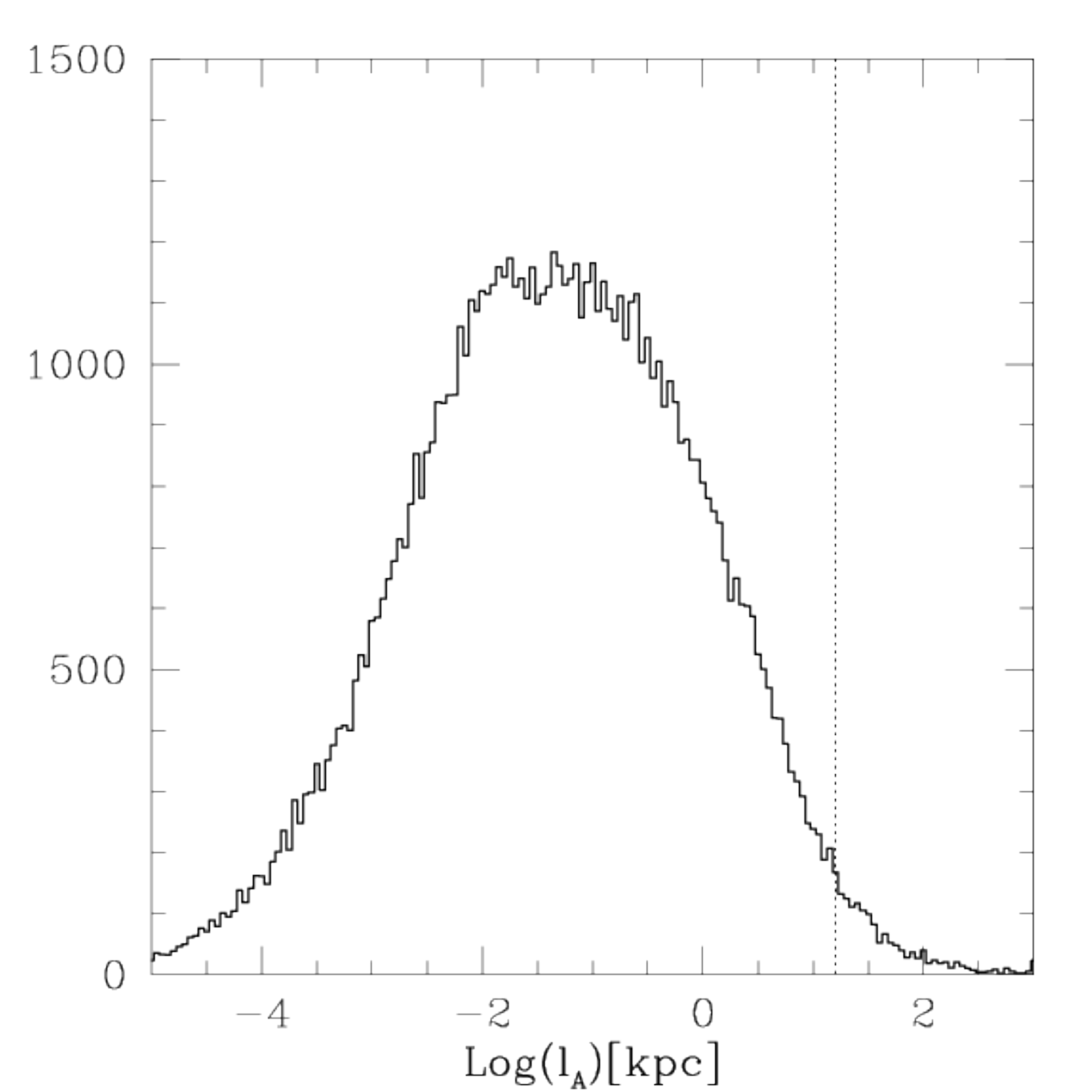}
\includegraphics[width=0.47\textwidth]{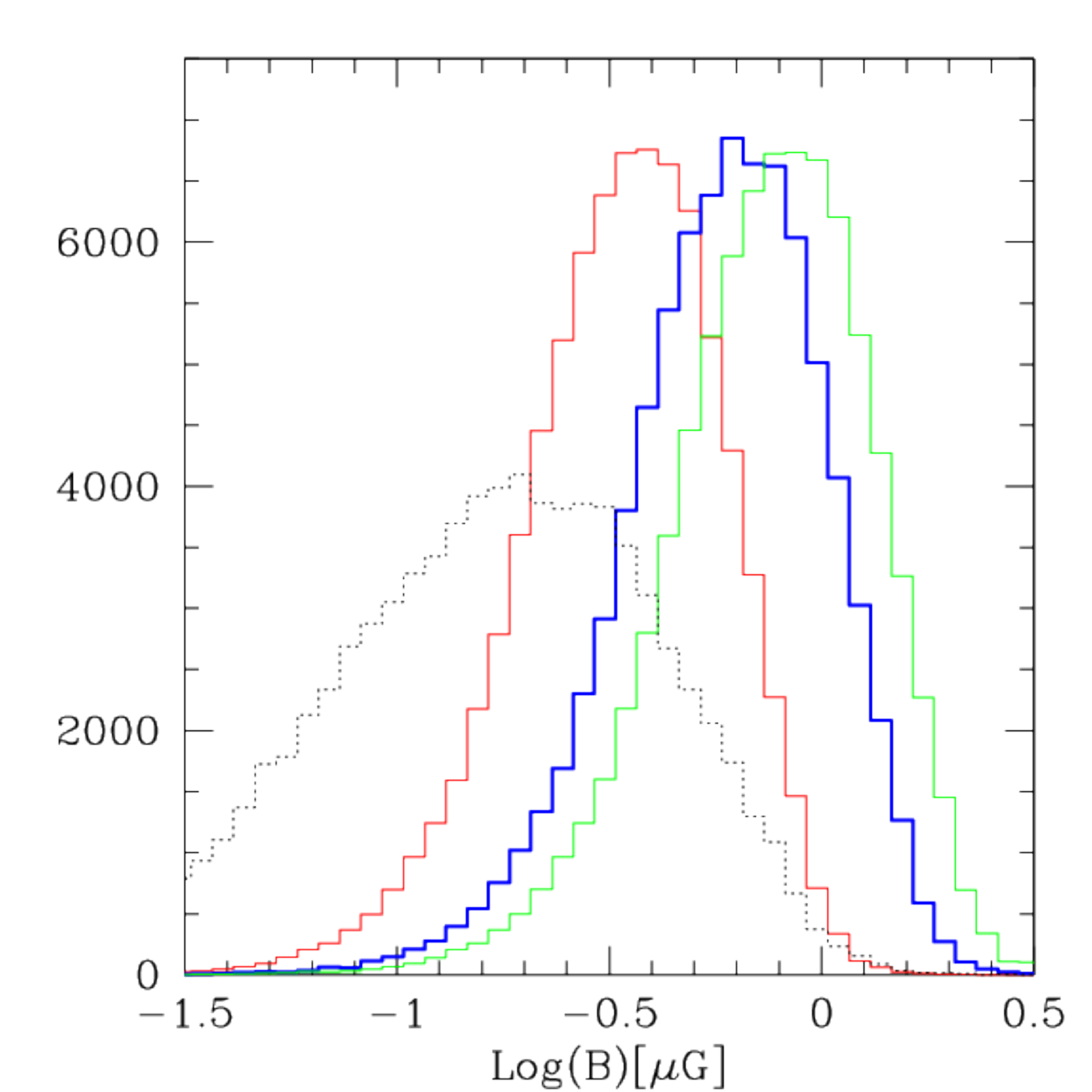}
\caption{Upper : Number-cell distribution as a function of the MHD scale. The vertical dotted line marks 16 kpc. 
Lower : Number-cell distribution of magnetic field intensity assuming $\eta_B=$ 0.02 (red), 0.03 (blue), 0.05 (green). Dotted-line histogram shows the distribution of the original magnetic field measured in the simulation.}
\label{Fig:lAandB}
\end{figure}

The Alfv\'{e}n scale $l_A$ is the minimum scale where turbulent Reynolds stresses can bend field lines, i.e. essentially where the turbulent velocity matches the Alfv\'{e}n speed.
For a Kolmogorov scaling this is:

\begin{equation}
l_A = L M_A^{-3} = {{ ({6 \over 5})^3 L}\over{
(\sqrt{\beta_{pl}} M)^3}}
\end{equation}

where $M=\delta V/c_s$ is measured at that scale $L$ that in simulations is the scale which is iteratively found by our algorithm \citep[][]{va17turb}. 
In Fig.\ref{Fig:lAandB} (top panel) we show the number distribution of 81000 cells (Fig.\ref{Fig:dens}) as a function of the MHD scale, compared with 16 kpc (our resolution is typically between 8 and 16 kpc in the bridge).
We find that the typical $l_A \sim 0.1$ kpc and that the MHD scale is smaller than the numerical resolution in 90 percent of the cells there.
Under these conditions, the simulated dynamo amplification is suppressed by numerical resolution \citep[e.g.,][]{donnertrev18} and consequently the magnetic field in our simulations should be considered a lower limit.

\noindent
In order to overcome the numerical limitations discussed above, we derive the magnetic field by adopting an a-posteriori model.
The turbulent spectrum in the bridges develops in about one eddy-turnover time, $L/\delta V \sim$ few 100 Myrs. As soon as turbulence reaches the dissipation scale a fraction of the energy flux of the solenoidal motions is converted into
magnetic fields.
Motivated by simulations of MHD turbulence \citep[][]{beresnyak12}, we assume that the amplification initially operates in a kinematic regime, where the magnetic field grows exponentially with time, $B^2(t) \sim B^2_0 \exp (t \, \Gamma)$, where the time-scale of the magnetic growth is $\Gamma^{-1} \sim 30 \tau_e /\sqrt{Re}$,
$\tau_e$ is the eddy turnover-time, $\tau_e \sim L/\delta V$. When the magnetic and the kinetic energy densities become comparable at the viscous dissipation scale,
the turbulent dynamo transits to a phase  where the magnetic field energy grows linearly with time. 
This transition occurs after a time:
\begin{equation}
\Delta T \sim 60 \tau_e {\rm Re}^{-1/2} \ln  
\left ( \frac{\sqrt{4 \pi \rho_{ICM}} \delta V}{B_0 {\rm Re}^{1/4} } \right )
\label{timeexp}
\end{equation}
that assuming the relevant parameters of the ICM in the A399-A401 bridge and $B_0 \sim 0.1 \mu$G (the average field measured in cosmological simulations) becomes shorter than a eddy turnover time as soon as ${\rm Re} > 10^3$.
\\
The ICM in the intra-cluster bridges shares conditions similar to the medium in the outskirts of galaxy clusters, $n_{ICM} \geq 10^{-4}$ cm$^{-3}$ and $T \sim$ 3-5 keV. 
In the presence of a magnetic field $B >> $nG, this medium  
is weakly collisional and unstable high beta plasma with the consequences that the reduced mfp due to instabilities make
the effective Reynolds number very large \citep[e.g.,][]{sche06,lazarianberesnyak06,instabilities10,brunettilazarian11mfp,santos14}.
For these reasons in the paper we have assumed that the exponential phase is fast enough so that the magnetic
field energy simply grows linearly with time for most of its evolution. 
As an additional approximation we also neglect the delay in the field amplification due to the turbulent cascading time and simply assume that $B^2/8\pi \sim \eta_B F \tau_e$, where $\eta_B =$ 0.02, 0.03 and 0.05.
A comparison between the original magnetic field strength  in simulations and the results from our simple modelling is shown in Fig.\ref{Fig:lAandB} (bottom panel) demonstrating that the effect of the dynamo from the conversion of the turbulent flux into magnetic fields is expected to be dominant.
In principle, the amplification should be calculated in post processing following the evolution with time of the turbulence in the simulation. However, dynamically active bridges are short lived systems, with a life-time constrained by the bridge crossing time $\tau \sim R/V_i$, that is $\sim 1-1.5$ Gyr considering an intra-cluster
impact velocity $V_i \sim 2000-3000$ km s$^{-1}$ and a bridge length $R \sim 3$ Mpc. This life-time is similar to (only slightly larger than) the typical eddy turnover time of
turbulence measured in simulations, $\tau_e \sim L/\delta V \sim 0.4-1$ Gyr. 
As a consequence assuming that the magnetic field is amplified within about 1 eddy 
turnover time provides a decent approximation here.

\section{Comparison with TTD acceleration}

\begin{figure}
\centering
\includegraphics[width=0.47\textwidth]{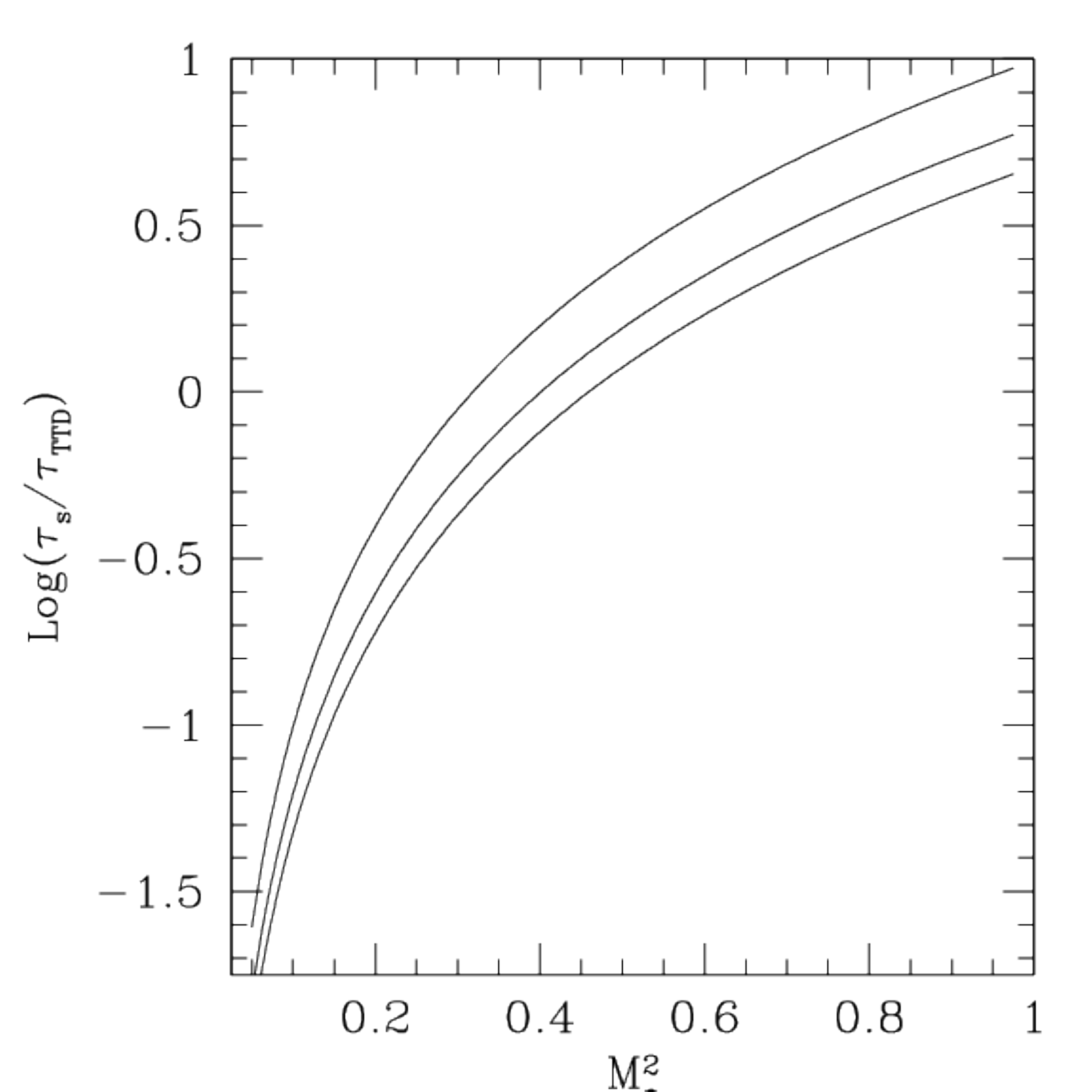}
\caption{Ratio of incompressible and TTD turbulent acceleration times as a function of turbulent (compressive) Mach number. We assume $\eta_B=0.3$, $L_s/L_c =1$ and $c_s=10^8$ km s$^{-1}$.
The three lines are obtained assuming $F_s/F_c$=1, 2, 3 (from top to bottom).}
\label{Fig:ratio2}
\end{figure}

In this paper we have assumed that relativistic particles are re-accelerated and decelerated in a systematic way in reconnecting and magnetic-dynamo regions, respectively, and on longer time-scales undergo a stochastic second order Fermi process diffusing across these sites in super-Alfv\'enic MHD turbulence \citep[][]{2016MNRAS.458.2584B}. This mechanisms, proposed for the ICM, was applied also to gamma-ray bursts and Pulsars Wind Nebulae \citep[][]{xu17, xu19}.
The diffusion coefficient in the particles momentum space induced by this mechanism is \citep[][]{2016MNRAS.458.2584B}: 

\begin{equation}
    D_{pp} \sim \left({{l_A}\over{\lambda_{mfp}}} \right)^2 
    {{V_A^2}\over{D}} p^2
    \label{dpp1}
\end{equation}

where $\lambda_{mfp}$ is the effective particles mfp and $D \sim 1/3 c \lambda_{mfp}$ is the spatial diffusion coefficient. 
In super-Alfv\'enic turbulence hydro motions set $\lambda_{mfp} \leq l_A$, because particles travelling in magnetic fields tangled on scales $\geq l_A$ change directions on this scale preserving the adiabatic invariant. 
In addition magnetic field fluctuations in MHD turbulence induce resonant interaction with particles and pitch-angle scattering with respect to the local field direction. In super-Alfv\'enic turbulence the interaction is
driven by the largest moving mirrors on scales $L \sim l_A$ and - similar to magnetic field tangling - limits the
effective mfp to $\lambda_{mfp} \leq l_A$ (see discussion in \citep[][]{2016MNRAS.458.2584B}).
Following \citep[][]{2016MNRAS.458.2584B} we assume a situation where kinetic effects on smaller scales are sub-dominant for relativistic particles and adopt a value of the effective $\lambda_{mpf}$ that is a fraction of (similar to) $l_A$, specifically $\lambda_{mfp} \approx 1/2 l_A$ \citep[e.g.,][]{2016MNRAS.458.2584B}. This gives the diffusion 
coefficient in the momentum space (from eq.\ref{dpp1}) :

\begin{equation}
D_{pp} \simeq
{{48}\over c} {{F}\over{\rho_{ICM} V_A}}
p^2 \left( {{\psi}\over{1/2}} \right)^{-3}
\label{dpp2}
\end{equation}

\noindent
that is adopted in the paper.
\\
TTD acceleration in compressive MHD turbulence is the mechanism that is typically assumed in galaxy clusters to calculate radio halo models \citep[e.g.,][]{brunettilazarian07,pinzke17}. 
It has been also claimed that the efficiency of this mechanisms in the ICM depends on the effective collisionality of the plasma, being stronger in the case of reduced effective mfp of the thermal ICM \citep[][]{brunettilazarian11mfp}. 
In this Section we focus on the collisionless version of the TTD mechanism, and compare the efficiency of this mechanism with that of the stochastic acceleration from solenoidal super-Alfv\'{e}nic turbulence that we have adopted in this paper.
Combining eqs.8-9 in \citep{brunetti16} with eq.\ref{dpp2} and considering $\tau \sim p^2/(4 D_{pp})$, we find :

\begin{equation}
{{\tau}\over{\tau_{TTD}}} \sim 14 \tilde{f} ({{\psi}\over{1/2}})^3
\eta_B^{1 \over 2} {{F_c}\over{F_s}} {{M_c}\over{\sqrt{\beta_{pl}}}}
\label{ratio1}
\end{equation}

where $F_c$ and $F_s$ are the kinetic energy fluxes of the compressive (fast modes)
and solenoidal turbulence, respectively, $M_c$ is the turbulent Mach number of the compressive
turbulence, 
\begin{equation}
\tilde{f}=x^4+x^2-(1+2 x^2)\ln(x)-{5 \over {4}}
\label{fff}
\end{equation}
and $x=c_s/c$ ($c_s$ is the sound speed in the ICM).
Eq.\ref{ratio1} means that, in the case of significant solenoidal component and for $M/\sqrt{\beta_{pl}}<<1$, the acceleration by incompressible turbulence may become faster than TTD. We note that in the case of subsonic turbulence the above condition implies, that
for strongly super-Alfv\'enic tubulence, ie $M^2 \beta_{pl}>>1$, the acceleration rate due to incompressible motions may be larger than TTD.
If $\beta_{pl}$ is derived from our simple model of magnetic field eq.\ref{ratio1} is :
\begin{equation}
    {{\tau}\over{\tau_{TTD}}} \sim 12.8 \tilde{f} ({{\psi}\over{1/2}})^3
\eta_B^{1 \over 2} ({{F_c}\over{F_s}})^{2 \over 3} M_c^2
\sqrt{{{L_s}\over{L_c}}}
\label{ratio2}
\end{equation}
In Fig.\ref{Fig:ratio2} we show the ratio of the acceleration times-scales (eq.\ref{ratio2}) as a function of the turbulent Mach number, assuming similar injection scales for the solenoidal and for the compressive components, i.e. $L_s/L_c \sim 1$ (see caption). We find that the two mechanisms have similar acceleration rates considering $M_c^2 \sim 0.2-0.5$, TTD is more efficient for larger Mach numbers. Specifically in our simulations we measure $F_s/F_c \sim 2.5$ in the bridge region and $M^2_c \sim 0.2-0.3$ implying that TTD is slightly subdominant, although it can provide an additional contribution.

\end{document}